\documentclass[12pt,reqno]{article}
\usepackage{amsmath}
\usepackage{amsthm}
\usepackage{amsfonts}

\input{xy}
\xyoption{all}

\newcommand{\oh}{{\scriptstyle\frac{1}{2}}}

\newcommand{\zig}{{\mathcal{Z}}}
\newcommand{\azig}{{\mathcal{\bar{Z}}}}
\newcommand{\lla}{{\ \longleftarrow\ }}
\newcommand{\lra}{{\ \longrightarrow\ }}
\newcommand{\gl}{{\mathfrak{gl}}}
\newcommand{\sll}{{\mathfrak{sl}}}
\newcommand{\psl}{{\mathfrak{psl}}}
\newcommand{\Span}{{\text{span}}}

\DeclareMathOperator{\sign}{\text{sign}}
\DeclareMathOperator{\End}{\text{End}}

\newcommand{\Complex}{\mathbb{C}}

\newcommand{\Integer}{\mathbb{Z}}
\newcommand{\mat}{\begin{pmatrix}}
\newcommand{\tam}{\end{pmatrix}}
\newcommand{\smat}{\left(\begin{smallmatrix}}
\newcommand{\mc}{\mathcal}
\newcommand{\mf}{\mathfrak}

\def\g{\mathfrak{g}}
\def\h{\mathfrak{h}}

\def\sfrac12{{\scriptstyle \frac12}}
\def\P{{\cal P}}
\def\S{{\cal S}}

\title{\bf Representation theory of $\sll(2|1)$}  
\author{\\[5mm] Gerhard G\"otz$^1$, Thomas Quella$^2$, 
               Volker Schomerus$^1$ \\[5mm] 
$^1$ Service de Physique Th\'eorique, CEA Saclay,\\ 
F-91191 Gif-sur-Yvette, France\\[5mm] 
$^2$ King's College London, Department of Mathematics, \\ 
Strand, London WC2R 2LS, UK\\[5mm]}
\date{}
\begin{document}
\begin{titlepage}      \maketitle       \thispagestyle{empty}

\vskip1cm
\begin{abstract} 
In this note we present a complete analysis of finite dimensional 
representations of the Lie superalgebra $\sll(2|1)$. This includes, 
in particular, the decomposition of all tensor products into their
indecomposable building blocks. Our derivation makes use of a close 
relation with the representation theory of $\gl(1|1)$ for which 
analogous results are described and derived. 
\end{abstract} 

\vspace*{-15.9cm}\noindent 
{\tt {hep-th/0504234}} \\
{\tt {KCL-MTH-05-04}} \\
{\tt {SPhT-T05/062}} \\ 
\bigskip\vfill
\noindent

{\small e-mail: }{\small\tt
Gerhard.Goetz@cea.fr \& quella@mth.kcl.ac.uk \& 
Volker.Schomerus@cea.fr } 
\end{titlepage} 

\baselineskip=19pt 
\setcounter{equation}{0} 
\section{Introduction}

Since their first systematic discussion \cite{Kac:1977em} in the 
1970's, Lie superalgebras have been studied for a variety of
reasons, both in physics and in mathematics. They found applications
not only in elementary particle physics (see \cite{Nilles:1983ge} 
for an early review) but also to condensed matter problems, mostly 
in the context of disordered fermions \cite{Efetov1983:MR708812}
and in particular the quantum Hall effect \cite{Gruzberg:1999dk,
Zirnbauer:1999ua} (see also e.g.\ \cite{Read:2001pz} for further 
applications to models of statistical physics). During the last 
years, non-linear $\sigma$-models on supergroups and supercosets 
have also emerged through studies of string theory in certain 
Ramond-Ramond backgrounds \cite{Metsaev:1998it,Rahmfeld:1998zn,
Berkovits:1999im}. Many special properties of these models, such 
as the possible presence of conformal invariance without a 
Wess-Zumino term, originate from peculiar features of the underlying 
Lie superalgebra \cite{Bershadsky:1999hk,Berkovits:1999zq}.
\smallskip

Even though Lie superalgebras are so widely used, their representation 
theory, and in particular their Clebsch-Gordan decomposition, is far 
from being fully developed. This may partly be explained by 
the fact that indecomposable (but reducible) representations occur 
quite naturally \cite{Kac:1977em,Scheunert:1977wj,Marcu:1979se}. 
Furthermore, many Lie superalgebras are known not to admit a 
complete classification of all finite dimensional representations 
\cite{Germoni1998:MR1659915}. One of the rare exceptions for which 
such a classification exists are the Lie superalgebras of type 
$\sll(n|1)$ \cite{Shmelev1982:MR687395,Leites1982:MR683439,
Germoni1998:MR1659915,Su2000:MR1738588}.%
\smallskip

In this note we shall discuss the representation theory of $\sll(2|1)$, 
including a {\em complete} list of tensor products of finite dimensional 
representations with diagonalizable Cartan elements. Thereby, we extend 
previous partial studies \cite{Scheunert:1977wj,Marcu:1979sg}. Our 
derivations are based on a particular embedding of the Lie superalgebra 
$\gl(1|1)$. For the purpose of being self-contained we shall therefore 
commence in section \ref{sc:GL} with a short exposition of the Lie 
superalgebra $\gl(1|1)$, its finite dimensional representations and 
their tensor products. 
\smallskip

All our new results on $\sll(2|1)$ are contained in section \ref{sc:TP}. 
First, we investigate how $\sll(2|1)$ representations decompose after 
restricting the action to the subalgebra $\gl(1|1)$. These decompositions 
exhibit a very close correspondence between atypical representations 
(short multiplets) of $\gl(1|1)$ and $\sll(2|1)$. The latter extends 
to indecomposable composites of atypical representations. Our results 
for the decomposition of $\sll(2|1)$ tensor products into their 
indecomposable building blocks are stated in the propositions 1, 2 
and 4. Proposition 3 states that, modulo projectives, the representation 
ring of $\sll(2|1)$ may be embedded into the representation ring of
$\gl(1|1)$.%
\smallskip

In a forthcoming publication \cite{Gotz:2005ka} we shall employ the 
results of this paper and related ideas in order to determine the 
tensor products of a large class of $\psl(2|2)$-representations.
The latter are relevant for the study of strings in $AdS_3$. In 
addition, our analysis might possess implications for the 
construction of new conformal fields theories with $\gl(1|1)$ 
or $\sll(2|1)$ superalgebra symmetries (see, e.g.,\ 
\cite{Rozansky:1992rx,Maassarani:1996jn,Essler:2005ag}). As in the case of 
bosonic current algebras, the representation theory of affine
Lie superalgebras inherits much of its features from the finite 
dimensional algebra of zero modes. But in the case of current 
superalgebras, there remain many unresolved issues, e.g.\ concerning the 
modular transformation of characters and the relation of the modular 
$S$ matrix to the fusion algebra \cite{Rozansky:1992td,Semikhatov:2003uc,
Jacobsen:2005uw}. We hope to come back to these important questions
in the future. Let us finally also note that $\gl(1|1)$ symmetry 
has been argued to be an imminent feature of every $c=0$ 
conformal field theory \cite{Gurarie:1999yx,Gurarie:2004ce}.%

\section{\label{sc:GL}The Lie superalgebra $\gl(1|1)$}
\setcounter{equation}{0} 

This section is devoted to the representation theory of $\gl(1|1)$.  
Not only will this Lie superalgebra play a crucial role when we 
determine tensor products of $\sll(2|1)$ representations, it can 
also serve as a very instructive example in which we encounter 
some of the most interesting phenomena and notions in the 
representation theory of Lie superalgebras.

\subsection{The defining relations}

The Lie superalgebra $\h= \gl(1|1)$ is generated by two even elements 
$E$, $N$ and two odd elements $\psi^\pm$ (we shall follow the physicists 
convention of \cite{Rozansky:1992rx}). The element $E$ is central and the 
fermions $\psi^\pm$ have opposite charge with respect to $N$. More 
explicitly the defining relations read,  
\begin{align}
  [E,\psi^\pm]
  &\ =\ [E,N]\ =\ 0&
  [N,\psi^\pm]
  &\ =\ \pm\psi^\pm&
  \{\psi^+,\psi^-\}
  &\ =\ E\ \ .
\end{align}
The even subalgebra is thus given by $\h^{(0)} = \gl(1)\oplus \gl(1)$. 
For later convenience let us also introduce the automorphism $\omega$ 
which is defined by its action 
\begin{align}
  \label{eq:AutoGL}
  \omega(E) & \ = \ E & \omega(N) & \ = \ - N & \omega(\psi^\pm)
  & \ =\ \psi^\mp
\end{align}
on the generators and extended to the whole Lie superalgebra $\h$ by 
linearity.%

\subsection{The finite dimensional representations}

The indecomposable finite dimensional representations of $\h=\gl(1|1)$ have 
been classified in \cite{Kac:1977em,Kac1977:MR0444725,Leites1982:MR683439,
Su2000:MR1738588}.\footnote{To avoid confusion we stress that the term 
{\em indecomposable} refers to both irreducible as well as reducible 
but indecomposable representations.} We shall start with a short 
discussion of irreducible 
representations which may all be obtained from the so-called Kac modules. 
In this context it is  crucial to distinguish between {\em typical} and 
{\em atypical} representations \cite{Kac1978:MR519631}, or long and short 
multiplets. The most striking feature of the latter is that 
they can be part of larger indecomposable representations. A complete 
list of such ``composites'' is provided in the second and third 
subsection.

\subsubsection{Kac modules and irreducible representations}

Let us agree to work with a Cartan subalgebra that is spanned by $E$ and $N$. 
In order to introduce Kac modules we define $\psi^+$ to be a positive root
and $\psi^-$ to be a negative root. The Kac modules $\langle e,n\rangle$ are 
induced highest weight modules over a one-dimensional representation $(e,n)$ 
of the bosonic subalgebra, where $e\in\Complex$ and $n\in\Complex$ are the
eigenvalues of $E$ and $N$, respectively. A more explicit description 
through matrices is, 
\begin{align}
  \langle e,n\rangle:&&
  E&\ =\ \mat e&0\\0&e\tam&
  N&\ =\ \mat n&0\\0&n-1\tam&
  \psi^+&\ =\ \mat0&e\\0&0\tam&
  \psi^-&\ =\ \mat0&0\\1&0\tam \ \ . 
\end{align}
Similarly, one can introduce anti-Kac modules $\overline{\langle
e,n \rangle}$ by switching the role of positive and negative roots. The 
corresponding matrix representation reads, 
\begin{align}
  \overline{\langle e,n\rangle}:&&
  E&\ =\ \mat e&0\\0&e\tam&
  N&\ =\ \mat n-1&0\\0&n\tam&
  \psi^+&\ =\ \mat0&0\\1&0\tam&
  \psi^-&\ =\ \mat0&e\\0&0\tam\ \ .
\end{align}
Note that the modules $\langle e,n\rangle$ and $\overline{\langle e,n
\rangle}$ are irreducible if and only if $e\neq0$ in which case they 
are also isomorphic. The resulting representations are called typical
and they provide the ``generic'' irreducible representations of $\gl(1|1)$.  
\smallskip 

For $e=0$, on the other hand, one obtains two inequivalent 
indecomposable representations both of which possess an invariant 
one-dimensional subspace. If we introduce the notation $\langle n
\rangle$ for the one-dimensional irreducible representations 
specified by
\begin{align}
  E&\ =\ 0&
  N&\ =\ n&
  \psi^\pm&\ =\ 0
\end{align}
then we may express the structure of the indecomposable Kac and 
anti-Kac modules through the following diagrams, 
\begin{align}
  \langle 0,n\rangle:&\qquad
  \langle n-1\rangle\ \lla\ \langle n\rangle&
  \overline{\langle 0,n\rangle}:&\qquad
  \langle n-1\rangle\ \lra\ \langle n\rangle\ \ .
\end{align}
Pictures of this type (and certainly much more complicated versions) 
will appear frequently throughout this text. Let us therefore pause 
for a moment to recall how we decode their information: Atypical 
representations from which no arrows emanate correspond to invariant 
subspaces. If we divide by such a subrepresentation, the resulting 
quotient is encoded by a new diagram which is obtained from the 
original by deleting the invariant subspace along with all the 
adjacent arrows. In the case of (anti-) Kac modules there exists 
only a single irreducible invariant subrepresentation and the 
corresponding quotients are irreducible. But we will soon see 
examples of modules with several invariant subspaces or even 
whole hierarchies thereof. In such cases, our diagrams may have 
different floors which are connected by arrows.

\subsubsection{Projective covers of atypical irreducible representations}

We have observed already that the atypical irreducible representations
can be part of larger indecomposables, e.g.\ of the Kac and anti-Kac modules. 
The latter can themselves appear as proper submodules of indecomposable 
structures. There exist certain distinguished indecomposables, however, 
that admit no further extension. These are the so-called projective 
covers $\P_\h(n)$ of atypical representations that we are going to 
introduce next.    
\smallskip 

The representations $\P_\h(n)$  are four-dimensional and they  
are parametrized by one complex parameter $n$ which features 
explicitly in the following matrices, 
\begin{align}
  N&\ =\ \mat n&0&0&0\\0&n+1&0&0\\0&0&n-1&0\\0&0&0&n\tam&
  \psi^+&\ =\ \mat 0&0&0&0\\1&0&0&0\\0&0&0&0\\0&0&1&0\tam&
  \psi^-&\ =\ \mat 0&0&0&0\\0&0&0&0\\1&0&0&0\\0&-1&0&0\tam\ \ .
\nonumber
\end{align}
The element $E$ vanishes identically. It is worth mentioning that 
$\P_\h(0)$ is the adjoint representation of $\gl(1|1)$. As they 
stand, the matrices are not very illuminating. In fact, the 
structure of $\P_\h(n)$ is much better understood after  
translation into our diagrammatic language, 
\begin{equation}
 \P_\h(n):\qquad
  \langle n\rangle\ \lra\ \langle n+1\rangle\oplus\langle n-1\rangle\ 
\lra\ \langle n\rangle\ \ .
\end{equation}
There is a variant of this pictorial presentation that keeps track 
of the ordering of the weights, i.e.\ of the eigenvalues for $N$, 
\begin{equation}\label{Ppic2}
{\xymatrixrowsep{4pt}
\xymatrixcolsep{4pt}
  \xymatrix{ & \langle n\rangle\ar[dr]\ar[dl]&\\
     \langle n-1\rangle\ar[dr] && \langle n+1\rangle\ \ar[dl]
                     \ . \\
             &\langle n\rangle &}}
\end{equation}
In this diagram, $N$-eigenvalues increase from left to right. Both 
pictures display the essential features of $\P_\h(n)$ very clearly. 
To begin with, these representations contain a unique irreducible 
one-dimensional subrepresentation $\langle n \rangle$ in the rightmost
position (bottom). This is called the ``socle'' of $\P_\h(n)$ and it is 
the reason for us to think of the four-dimensional indecomposables 
as a ``cover'' of atypical irreducible representations. In addition, 
we can also find three different types of indecomposable 
subrepresentations in $\P_\h(n)$. These include the two-dimensional 
(anti-)Kac modules $\langle 0,n+1\rangle$ and $\overline{\langle 0,n
\rangle}$. But there appears also one new class of three-dimensional 
indecomposables that we did not meet 
before. Their diagram is obtained from the above by deleting the 
representation $\langle n \rangle$ on top along with the 
arrows that emanate from it. One can go through a similar 
analysis of factor representations obtained from $\P_\h(n)$
with very much the same pattern of results. Let us only 
point out that the quotient of $\P_\h(n)$ by its socle 
$\langle n\rangle$ provides a new three-dimensional 
indecomposable representation which is not isomorphic to 
the one we found among the submodules of $\P_\h(n)$. 
\smallskip 

We have seen that atypical irreducibles sit inside (anti-)Kac
modules which in turn appear as subrepresentations of three-%
dimensional indecomposables. But the sequence of embeddings
does not end here. In the next subsection we shall construct 
two infinite series of indecomposables which are  
nested into each other such that their $m^{th}$ member appears 
as an extension of the $(m-1)^{th}$ by a one-dimensional 
atypical representation. The representation $\P_\h(n)$ gives 
rise to another extension of three-dimensional indecomposables, 
but this one turns out to be maximal, i.e.\ no further 
embedding into a larger indecomposable is possible. Their 
maximality distinguishes $\P_\h(n)$ from all other 
representations with $E=0$ and it places them in one group 
with the typical two-dimensional representations. In more 
mathematical terms, $\langle e,n\rangle, e \neq 0,$ and 
$\P_\h(n)$ are known as {\em projective} representations 
of $\gl(1|1)$, a notion that is particularly important for
our investigation of tensor products since the projective 
representations form an ideal in the representation ring.

\subsubsection{Zigzag modules}
\def\Z{{\mc{Z}}}
\def\bZ{{\bar{\mc{Z}}}}

As we have anticipated at the end of the previous subsection, there exist 
two different families of indecomposable representations $Z^d_\h(n)$ and 
$\bZ^d_\h(n)$ which we shall name (anti-) {\em zigzag} representations.
They are parametrized by the eigenvalue $n \in \mathbb{C}$ of $N$ with 
the largest real part and by the number $d= 1,2,3,\dots$ of their 
atypical constituents. 
On a basis of eigenstates $|m\rangle, m = n, \dots, n-d+1,$ for the 
element $N$, the generators of zigzag representations $\Z_\h^d(n)$ read   
\begin{equation} \label{zig} 
 N |m\rangle \ = \ m |m\rangle \ \ \ , \ \ \ 
    \psi^\pm |m\rangle \ = \ \frac12 \bigl(1 + (-1)^{n-m}\bigr)|m\pm1\rangle
\end{equation} 
and $E$ vanishes identically. Here we agree that $|m\rangle = 0$ when 
$m$ is outside the allowed range. Similarly, we can introduce anti-zigzag
representations $\bZ_\h^d(n)$ through      
\begin{equation} \label{azig}  
 \ \ \ N |m\rangle \ = \ m |m\rangle \ \ \ , \ \ \ 
    \psi^\pm |m\rangle \ = \ \frac12 \bigl(1 - (-1)^{n-m}\bigr)|m\pm1\rangle\ \ . 
\end{equation} 
The only difference between the formulas (\ref{zig}) and (\ref{azig}) is 
in the sign between the two terms for the action of fermionic elements. 
Note that atypical irreducible representations and (anti-)Kac modules are  
special cases of (anti-)zigzag representations, in particular 
we have $\langle n\rangle\cong\zig^1_\h(n)\cong \azig^1_\h(n)$. 
\smallskip 
 
Once more we can display the structure of the (anti-)zigzag modules 
through their associated diagram. In doing so we shall separate two 
cases depending on the parity of $d$. When $d=2p$ is even we find 
\begin{equation}
  \begin{split}\nonumber
    \mc{Z}^{2p}_\h(n):&\qquad\langle n-2p+1\rangle\lla
     \langle n-2p+2\rangle\lra\cdots\lla\langle n-2\rangle\lra
     \langle n-1\rangle\lla\langle n\rangle\\[2mm]
    \mc{\bar{Z}}^{2p}_\h(n):&\qquad\langle n-2p+1\rangle\lra
    \langle n-2p+2\rangle\lla\cdots\lra\langle n-2\rangle\lla
    \langle n-1\rangle\lra\langle n\rangle\ \ .
  \end{split}
\end{equation}
Observe that the leftmost atypical constituent is invariant for even 
dimensional zigzag modules, a property that is not shared by the even 
dimensional anti-zigzag representations which, by construction, always
possess an invariant constituent in their rightmost position. When $d
=2p+1$ is odd, on the other hand, the corresponding diagrams read 
\begin{equation}
  \begin{split}\nonumber
    \mc{Z}^{2p+1}_\h(n):&\qquad\langle n-2p\rangle\lra
    \langle n-2p+1\rangle\lla \cdots\lla\langle n-2\rangle\lra
    \langle n-1\rangle\lla \langle n\rangle\\[2mm]
    \mc{\bar{Z}}^{2p+1}_\h(n):&\qquad\langle n-2p\rangle\lla
    \langle n-2p+1\rangle\lra\cdots\lra\langle n-2\rangle\lla
    \langle n-1\rangle\lra\langle n\rangle\ \ . 
  \end{split}
\end{equation}
In this case, both ends of the anti-zigzag modules correspond to 
invariant subspaces. Tensor products of (anti-)zigzag representations
will turn out to depend very strongly on the parity of $d$. In analogy 
with our second graphical presentation \eqref{Ppic2} for the projective 
representations $\P_\h(n)$, one may be tempted to change our diagrams 
for $\Z$ and $\bZ$ a little bit by moving the sources up such that all 
arrows run at a $45$ degree angle. The resulting pictures explain our 
name ``zigzag module''.

\subsubsection{Action of the automorphism on modules}

When calculating the tensor products of $\gl(1|1)$ representations 
we can save some work by using the additional information that is 
encoded in the existence of the outer automorphism $\omega$. In fact, 
given any representation $\mu$ with map $\rho_\mu:\mf{g}\to\End(V)$ 
and an automorphism $\omega$, we may define the new representation 
$\omega(\mu)$ on the same space $V$ through the prescription 
$\rho_{\omega(\mu)}=\rho\circ\omega$. Depending on the choice of 
$\mu$, the new representation $\omega(\mu)$ will often turn out 
to be inequivalent to $\mu$.  
\smallskip

Let us briefly work out how the various representations of 
$\gl(1|1)$ are mapped onto each other. For the projective 
representations one easily finds 
\begin{align}
  \omega\bigl(\langle e,n\rangle\bigr)&\ =\ 
   \overline{\langle e,1-n\rangle}&
  \omega\bigl(\P_\h(n)\bigr)&\ =\ \P_\h(-n)\ \ .
\end{align}
The second assignment is easily found from the structure 
\eqref{Ppic2} of the projective cover along with the obvious
rule $\omega(\langle n\rangle)=\langle-n\rangle$. A 
similar argument also determines the action of the 
automorphism $\omega$ on zigzag representations,  
\begin{align}
  \omega\bigl(\mc{Z}^d_\h(n)\bigr)
  &\ =\ \begin{cases}
          \ \ \mc{\bar{Z}}^d_\h(d-n-1)&\text{ for }d\text{ even}\\[2mm]
          \ \ \mc{Z}^d_\h(d-n-1)&\text{ for }d\text{ odd}\ \ . 
        \end{cases} 
\end{align}
For anti-zigzag representations the same rules apply with 
the roles of $\Z$ and $\bZ$ being switched. What makes these 
simple observations useful for us is the fact that the fusion 
of representation respects the action of $\omega$. In other 
words, if $\mu_3$ is a subrepresentation of $\mu_1 \otimes 
\mu_2$, then $\omega(\mu_3)$ arises in the tensor product 
of $\omega(\mu_1)$ and $\omega(\mu_2)$ and their multiplicities
coincide.

\subsection{Decomposition of $\gl(1|1)$ tensor products}

We are now ready to spell out the various tensor products of finite  
dimensional representations of $\gl(1|1)$. Obviously, there are quite
a few cases to consider. For the tensor product of two typical 
representations one finds  
\begin{equation}
  \langle e_1,n_1\rangle\otimes\langle e_2,n_2\rangle\ = \ 
\begin{cases} 
\ \P_h(n_1+n_2-1)  &\text{ for }e_1+e_2=0\\[2mm]
\ \bigoplus_{p=0}^{1} \ \langle e_1+e_2,n_1+n_2-p\rangle
&\text{ for }e_1+e_2\neq0\ \ .
       \end{cases}
\end{equation}
This formula should only be used when $e_1,e_2 \neq 0$. Tensor products 
between atypical Kac modules will appear as a special case below when 
we discuss the multiplication of zigzag representations.   
\smallskip 

Next we would like to consider the tensor products involving projective 
covers $\P_\h$ in addition to typical representations. These are given by 
\begin{equation}
  \begin{split}
  \langle e,n\rangle\otimes\P_\h(m)
  &\ =\   \langle e,n+m+1 \rangle \oplus \, 2 \cdot 
   \langle e, n+m\rangle \oplus \, \langle e,n+m-1\rangle 
\\[2mm] 
  \P_\h(n)\otimes\P_\h(m)
  &\ =\ \P_\h(n+m+1)\oplus \, 2 \cdot \P_\h(n+m)\ \oplus  
   \, \P_\h(n+m-1)\ \ , \label{hPP} 
  \end{split}
\end{equation}
where we assume once more that $e\neq 0$ in the first line. We observe 
that typical representations and projective covers close under tensor 
products, in perfect agreement with the general behavior of projective 
representations. 
\smallskip

Tensor products between the projective representations and (anti-)zigzag 
modules are also easy to spell out  
\begin{eqnarray}
    \langle e,n\rangle\otimes\zig_{d}(m)
    & = &  \langle e,n\rangle\otimes\azig_{d}(m)\ = \ 
   \bigoplus_{p=0}^{d-1} \ \langle e,n+m-p \rangle \\[2mm]  
   \P_\h(n) \otimes\zig_{d}(m) & = & \P_\h(n) \otimes\azig_{d}(m) 
   \ = \ \bigoplus_{p=0}^{d-1} \  \P_\h(n+m-p)\ \ .  \label{hPZ} 
\end{eqnarray}
On the right hand side, only projective representations appear. We 
conclude that the latter form an ideal in the representation ring, 
just as predicted by general results in the theory of Lie 
superalgebras.  
\smallskip  

\def\nn{\nonumber}
The description of tensor products between (anti-)zigzag representations
requires the most efforts since we have to treat various cases separately, 
depending on the parity of the parameter $d$. 
\begin{equation}
\begin{split} 
       \zig^{2p_1}_\h(n_1)\otimes\azig^{2p_2}_\h(n_2)
    &\ =\ \bigoplus_{\nu_1=0}^{p_1-1}\bigoplus_{\nu_2=0}^{p_2-1}
         \, \P_\h(n_1+n_2-2\nu_1-2\nu_2-1)\ \ ,  \\[2mm] 
     \zig^{2p_1}_\h(n_1)\otimes\zig^{2p_2}_\h(n_2)
    &\ =\ \zig^{2p_1}_\h (n_1+n_2) \oplus \zig^{2p_1}_\h (n_1+n_2-2p_2+1)
         \  \oplus \ \nn \\[2mm]
    &\ \hspace*{2.2cm}  
       \bigoplus_{\nu_1=0}^{p_1-1}\bigoplus_{\nu_2=1}^{p_2-1}
         \, \P_\h(n_1+n_2-2\nu_1-2\nu_2)\ \ \text{ for } \ \ 
          p_1 \leq p_2\ \  , 
\end{split}\end{equation} 
\begin{equation} \begin{split} 
     \zig^{2p_1+1}_\h(n_1)\otimes\zig^{2p_2+1}_\h(n_2)
    &\ =\ \zig^{2(p_1+p_2)+1}_\h (n_1+n_2)\ \oplus \ 
        \bigoplus_{\nu_1=0}^{p_1-1}\bigoplus_{\nu_2=1}^{p_2}
         \, \P_\h(n_1+n_2-2\nu_1-2\nu_2)\ , \nn \\[2mm]
      \zig^{2p_1+1}_\h(n_1)\otimes\azig^{2p_2+1}_\h(n_2)
    &\ =\ \azig^{2(p_2-p_1)+1}(n_1+n_2-2 p_1) \ \oplus \nn \\[2mm]
    & \ \hspace*{1.8cm}  
       \bigoplus_{\nu_1=0}^{p_1-1}\bigoplus_{\nu_2=0}^{p_2}
         \, \P_\h(n_1+n_2-2\nu_1-2\nu_2-1)\ \ \text{ for } 
       \ p_1\leq p_2\ ,  \nn \\[2mm]
       \zig^{2p_1+1}_\h(n_1)\otimes\zig^{2p_2}_\h(n_2)
    &\ =\ \zig^{2p_2}_\h(n_1+n_2) \ \oplus \ 
        \bigoplus_{\nu_1=0}^{p_1-1}\bigoplus_{\nu_2=1}^{p_2}
         \, \P_\h(n_1+n_2-2\nu_1-2\nu_2)\ \ . \nn 
 \end{split}
\end{equation}
The remaining formulas can either be obtained by applying the outer 
automorphism $\omega$ to the ones we have displayed or by a formal 
conjugation of the above expression in which we replace $\zig$ by 
$\azig$ (and vice versa) while touching neither their arguments
nor the projective part at all.\footnote{Note that the described 
conjugation and the application of $\omega$ are two different 
operations.} Though we have not found these 
tensor products in the literature, we would not be surprised if 
they were known before. In any case, they may be derived by an 
explicit construction of the vectors that span the corresponding 
invariant subspaces in each tensor product. Let us also 
point out that the representation ring of $\gl(1|1)$ possesses 
many different subrings, i.e.\ there exist many different subsets
of representations which close under tensor products. We observe, 
for example, that (anti-)zigzag modules of any given even length 
(or even a finite set thereof) can be combined with projective 
representations to form an ideal in the fusion ring. 

\section{\label{sc:SL}The Lie superalgebra $\sll(2|1)$}
\setcounter{equation}{0}

This section is devoted to our main theme, the theory of finite  
dimensional representations of $\sll(2|1)$. The latter have been 
entirely classified \cite{Marcu:1979se,Shmelev1982:MR687395,Leites1982:MR683439,
Su2000:MR1738588}. This distinguishes $\sll(2|1)$ from most other 
members of the A-series of Lie superalgebras for which a 
classification is even known to be impossible \cite{Germoni1998:%
MR1659915}. Here we shall provide a complete list of tensor 
products of finite dimensional representations of $\sll(2|1)$, 
thereby extending previous partial results by Marcu \cite{Marcu:1979sg}. 
We shall achieve this with the help of a nice correspondence 
between the indecomposables of $\sll(2|1)$ and $\gl(1|1)$ which 
allows us to employ the results of the previous section.%

\subsection{The defining relations}

The even part $\mf{g}^{(0)}=\gl(1) \oplus \sll(2)$ of the Lie
superalgebra $\mf{g}=\sll(2|1)$ is generated by four bosonic 
elements $H$, $E^\pm$ and $Z$ which obey the commutation relations
\begin{align} \label{sl211} 
  [H,E^\pm]&\ =\ \pm E^\pm\quad,&
  [E^+,E^-]&\ =\ 2H\quad,&
  [Z,E^\pm]&\ =\ [Z,H]\ =\ 0\ \ .
\end{align}
In addition, there exist two fermionic multiplets $(F^+,F^-)$ and 
$(\bar{F}^+,\bar{F}^-)$ which generate the odd part $\mf{g}^{(1)}$.
They transform as $(\pm \oh,\oh)$ with respect to the even subalgebra, 
i.e.\ 
\begin{align} \label{sl212} 
  [H,F^\pm]&\ =\ \pm\frac{1}{2}F^\pm&
  [H,\bar{F}^\pm]&\ =\ \pm\frac{1}{2}\bar{F}^\pm\nn\\[2mm]
  [E^\pm,F^\pm]&\ =\ [E^\pm,\bar{F}^\pm]\ =\ 0&
  [E^\pm,F^\mp]&\ =\ -F^\pm&
  [E^\pm,\bar{F}^\mp]&\ =\ \bar{F}^\pm\\[2mm]
  [Z,F^\pm]&\ =\ \frac{1}{2}F^\pm&
  [Z,\bar{F}^\pm]&\ =\ -\frac{1}{2}\bar{F}^\pm\ \ .\nn
\end{align}
Finally, the fermionic elements possess the following simple 
anti-commutation relations
\begin{align} \label{sl213}
  \{F^\pm,F^\mp\}&\ =\ \{\bar{F}^\pm,\bar{F}^\mp\}\ =\ 0&
  \{F^\pm,\bar{F}^\pm\}&\ =\ E^\pm&
  \{F^\pm,\bar{F}^\mp\}&\ =\ Z\mp H
\end{align}
among each other. Formulas \eqref{sl211} to \eqref{sl213} provide 
a complete list of relations in the Lie superalgebra $\sll(2|1)$. 
\smallskip 

There are two different decompositions of $\sll(2|1)$ that 
shall play some role in our analysis below. One of them is 
the following triangular decomposition 
\begin{equation}
  \mf{g}\ =\ \mf{g}_+\oplus\mf{p}\oplus\mf{g}_-\ \ ,
\end{equation}
in which the Cartan subalgebra is given by $\mf{p}=\Span(H,Z)$, 
the positive roots span $\mf{g}_+=\Span(E^+,F^\pm)$ and the 
negative roots generate the third subspace $\mf{g}_-=\Span(E^-,
\bar{F}^\pm)$. This decomposition corresponds to a particular 
choice of the root system. Let us recall that for Lie 
superalgebras, the latter is not unique. 
\smallskip 

Another natural decomposition is obtained by collecting all 
bosonic generators in one subspace while keeping the fermionic 
generators in two separate sets,  
\begin{equation}
  \mf{g}\ =\ \mf{g}_1^{(1)}\oplus\g^{(0)}\oplus\mf{g}_{-1}^{(1)}\ \ . 
\end{equation}
Here, $\mf{g}_1^{(1)}=\Span(F^\pm)$ and $\mf{g}_{-1}^{(1)}=
\Span(\bar{F}^\pm)$. By declaring elements of these three 
subspaces to possess grade $(-1,0,1)$, respectively, we can 
introduce an $\Integer$-grading in the universal enveloping 
algebra. Fermionic elements possess odd grades so that the 
new grading is consistent with the usual distinction between 
even and odd generators. 
\smallskip 

As in our discussion of $\gl(1|1)$ above, it will be useful for 
us to exploit the symmetries of $\sll(2|1)$. In this case, they 
are described by an outer automorphism that acts trivially on 
the generators $E^\pm$ and $H$ while exchanging the barred and 
unbarred fermionic elements and reversing the sign of $Z$, i.e.\ 
\begin{equation}
  \label{eq:AutoSL}
  \Omega:\quad(H,E^\pm,Z,F^\pm,\bar{F}^\pm)
  \ \mapsto\ (H,E^\pm,-Z,\bar{F}^\pm,F^\pm)\ \ . 
\end{equation}
The existence of this $\mathbb{Z}_2$-automorphism will allow us 
to determine several tensor products rather easily.

\subsection{Finite dimensional representations}

Since there are different notations floating around in the mathematics
\cite{Kac:1977em} and in the physics literature \cite{Scheunert:1977wj,%
Frappat:1996pb} we shall give a short account of the basic constructions 
of modules and how they are related. Our discussion is restricted to 
finite dimensional representations in which the Cartan subalgebra can 
be diagonalized. More general representations have been discussed in
\cite{Marcu:1979se,Su2000:MR1738588} and \cite{Su2001:MR1842414}. An 
overview over the representations considered in this paper is given
in table \ref{tab:RepsSL}.%

\subsubsection{Kac modules and irreducible representations}

The basic tool in the construction of irreducible representations are 
again the Kac modules \cite{Kac:1977em}. In the case of $\g=\sll(2|1)$, 
these form a 2-parameter family $\{b,j\}$ of $8j$-dimensional 
representations. We may induce them from the $2j$-dimensional 
representations $(b-\oh,j-\oh)$ of the bosonic subalgebra $\g^{(0)}$ 
by applying the generators in $\g_1^{(1)}$, i.e.\ the pair $F^\pm$ 
of fermionic elements. Our label $b\in\Complex$ denotes a 
$\gl(1)$-charge and spins of $\sll(2)$ are labeled by $j=\oh,1,
\dots$. To be more precise, we must first promote the representation 
space of the bosonic subalgebra to a $\mf{g}^{(0)}\oplus
\mf{g}_{-1}^{(1)}$-module by declaring that its vectors are annihilated
when we act with elements $\bar F^\pm$. Then we can set 
\begin{equation} \nn 
   \{b,j\}
  \ =\ \text{Ind}_{\mf{g}^{(0)}\oplus\mf{g}_{-1}^{(1)}}^{\mf{g}}
                  V_{(b-\oh,j-\oh)}
  \ =\ \mc{U}(\mf{g})\otimes_{\mc{U}(\mf{g}^{(0)}\oplus\mf{g}_{-1}^{(1)})}
    V_{(b-\oh,j-\oh)}\ \ .
\end{equation}
In this formula, $\mc{U}(\mf{g})$ denotes the universal enveloping 
algebra of $\mf{g}$ and $V$ is the $2j$-dimensional representation 
space of the bosonic subalgebra, or, to be more precise, of the 
extented algebra $\mf{g}^{(0)}\oplus \mf{g}_{-1}^{(1)}$. Let us 
emphasize that there is a relative shift in the labels between 
the representation $\{b,j\}$ of the Lie superalgebra and the 
corresponding bosonic representation $(b-\oh,j-\oh)$. The shift
guarantees that the highest eigenvalue of $H$ in the whole 
module is given by $j$ and it corresponds to the conventions 
of \cite{Scheunert:1977wj}. Even though the latter seem somewhat 
unnatural from the point of view of Kac modules we will later 
encounter some simplifications which justify this choice. 
\smallskip

The dual construction which promotes the fermions in $\g_{-1}^{(1)}$,
i.e.\ the generators $\bar{F}^\pm$, to creation operators yields 
anti-Kac modules ($b$ and $j$ take the same values as above)
\begin{equation}\nonumber 
  \overline{\{b,j\}}
  \ =\ \text{Ind}_{\mf{g}^{(0)}\oplus\mf{g}_1^{(1)}}^{\mf{g}}
       V_{(b+\oh,j-\oh)}
  \ =\ \mc{U}(\mf{g})\otimes_{\mc{U}(\mf{g}^{(0)}\oplus\mf{g}_1^{(1)})}
        V_{(b+\oh,j-\oh)}\ \ .
\end{equation}
This bosonic content of (anti-)Kac modules may be read off rather easily 
form their construction,
\begin{equation}
  \label{typdec}
  \begin{split}
    \{b,j\}\bigr|_{\mf{g}^{(0)}}
    &\ =\ (b-\oh,j-\oh)\ \otimes \ 
    \mc{U}(\g_1^{(1)})\bigr|_{\mf{g}^{(0)}}
  \\[2mm]
    \overline{\{b,j\}}\bigr|_{\mf{g}^{(0)}}
    &\ =\ (b+\oh,j-\oh)\ \otimes \ 
   \mc{U}(\g_{-1}^{(1)})\bigr|_{\mf{g}^{(0)}}
  \\[2mm] 
\text{ where } \ \ \ \ \ & \mc{U}(\g_{\pm1}^{(1)})\bigr|_{\mf{g}^{(0)}} \ = \ 
   \bigl[(0,0)\oplus(\pm\oh,\oh)\oplus(\pm1,0)\bigr]\ \ .
  \end{split}
\end{equation}
The product $\otimes$ on the right hand side denotes the tensor 
product of $\g^{(0)}$ representations. For generic values of $b$ 
and $j$, the modules $\{b,j\}$ and $\overline{\{b,j\}}$ are 
irreducible and isomorphic. At the points $\pm b=j$, however, 
they degenerate, i.e.\ the representations are indecomposable 
and no longer isomorphic. In fact, Kac and anti-Kac modules 
are then easily seen to possess different invariant 
subspaces.
\smallskip

By dividing out the maximal submodule from each Kac module $\{\pm j,j\}$ 
we obtain irreducible highest weight representation $\{j\}_\pm$ of 
dimension $4j+1$.\footnote{A similar construction using anti-Kac modules 
instead of Kac modules leads to the same set of representations.} 
In order to understand their structure in more detail, we 
emphasize that the representations $\{j\}_+$ with $j=0,\oh,\dots$ 
are constructed from the Kac modules $\{j+\oh,j+\oh\}$ by decoupling 
the states in the representation $(j+\oh,j+\oh)\oplus(j+1,j)$ of the 
bosonic subalgebra. For the representations $\{j\}_-$ with $j=\oh,1,
\dots$, on the other hand, we start from the Kac modules $\{-j,j\}$ 
and decouple the bosonic multiplets $(-j,j-1)$ and $(-j+\oh,j-\oh)$. 
This construction implies that the bosonic content of atypical 
representations is given by 
\begin{equation}
  \{j\}_\pm\ =\ \begin{cases}
                  (j,j)\oplus(j+\oh,j-\oh)&,\text{ for }+
             \text{ and }j=\oh,1,\dots\\[2mm]
                  (-j,j)\oplus\bigl(-(j+\oh),j-\oh\bigr)&,
            \text{ for }-\text{ and }j=\oh,1,\dots\ \ 
                \end{cases}
\end{equation}
and by $(0)$ in case of the trivial representation $\{0\} = \{0\}_+$.
Note that the representations $\{j\}_\pm$ are labeled by a non-negative 
$j$. From time to time we shall adopt a notation in which the label $\pm$ 
is traded for a sign in the argument, i.e.\ we set $\{l\} = 
\{|l|\}_{\sign(l)}$. In case of the trivial representation, this 
convention amounts to omitting the subscript $+$. The irreducible 
representations $\{b,j\}$ with $\pm b\neq j$ are called typical. 
All other irreducibles of the type $\{j\}_\pm$ are atypical. The 
$8$-dimensional adjoint representation is given by $\{0,1\}$, i.e.\ 
it is typical. 
\smallskip 

Let us note in passing that our outer automorphism $\Omega$ acts on 
the irreducible representations much in the same way as for $\gl(1|1)$
(see eq.\ \eqref{eq:AutoSL}). It is not difficult to see that 
\begin{align}
  \Omega\bigl(\{b,j\}\bigr)&\ =\ \{-b,j\}&
  \Omega\bigl(\{j\}_\pm\bigr)&\ =\ \{j\}_\mp\ \ . 
\end{align}
The second formula will be particularly useful to understand the 
action of $\Omega$ on indecomposable representation of $\gl(1|1)$. 
\smallskip 

As a byproduct of the construction of irreducible representations 
we have seen the first examples of indecomposables of $\sll(2|1)$, 
namely the (anti-)Kac modules $\{\pm j,j\}$ and $\overline{\{\pm j,j\}}$.
They are built from two atypical representations such that  
\begin{equation}
  \begin{split}
    \{\pm j,j\}:&\qquad\{j\}_\pm\ \longrightarrow\ \{j-\oh\}_\pm\\[2mm]
    \overline{\{\pm j,j\}}:&\qquad\{j-\oh\}_\pm\ \longrightarrow\ 
     \{j\}_\pm\ \ .
  \end{split}
\end{equation}
We shall construct many other indecomposables in the following 
subsections. Let us also note that Kac and anti-Kac modules are 
mapped onto each other by the action of our automorphism 
\eqref{eq:AutoSL}.%
\smallskip

We wish to stress that in the physics literature the construction
of representations originally proceeded along a different line
\cite{Scheunert:1977wj}. Here the existence of a state $|b,j\rangle$
with maximal $H$-eigenvalue $j$ (and $Z$-eigenvalue $b$) was assumed
on which $E^+$, $F^+$ and $\bar{F}^+$ acted as annihilators while
the generators $E^-$, $F^-$ and $\bar{F}^-$ have been used to construct
the remaining states. The shift in the definition of the Kac module
above is reminiscent of these different conventions. Note that
the summary on tensor products which can be found in 
\cite{Frappat:1996pb} uses the physical conventions of the 
original articles \cite{Marcu:1979se,Marcu:1979sg}.%

\subsubsection{Projective covers of atypical irreducible modules}

When we discussed the representations of $\gl(1|1)$ we have 
already talked about the concept of a projective cover of an 
atypical representation. By definition, the projective cover 
of a representation $\{j\}_\pm$ is the largest indecomposable 
representation $\mc{P}_\g^\pm(j)$ which has $\{j\}_\pm$ as a 
subrepresentation (its socle). We do not want to construct these 
representations explicitly here. Instead, we shall display how  
they are composed from atypicals. The projective cover of the 
trivial representation is an $8$-dimensional module of the form 
\begin{equation}
  \mc{P}_\g(0):\qquad\{0\}\ \longrightarrow\ \{\oh\}_+\oplus\{\oh\}_-\ 
    \longrightarrow\ \{0\}\ \ .
\end{equation}
For the other atypical representations $\{j\}_\pm$ with $j=\oh,1,\dots$
one finds the following diagram,  
\begin{equation} \label{P} 
  \mc{P}_\g^\pm(j):\qquad\{j\}_\pm\ \longrightarrow\ \{j+\oh\}_\pm
\oplus\{j-\oh\}_\pm\ \longrightarrow\ \{j\}_\pm\ \ .
\end{equation}
These representation spaces are $16j+4$-dimensional. A rather 
explicit constructions of the modules $\mc{P}_\g^\pm(j)$ with 
$j \neq 0$ will be sketched in the next section. Let us also 
agree to absorb the superscript $\pm$ on $\mc{P}$ into the 
argument, i.e.\ $\mc{P}_\g^\pm(j)= \mc{P}_\g(\pm j)$, wherever 
this is convenient.    

\subsubsection{Zigzag modules}

There are two additional sets of indecomposables that are close relatives
of the (anti-) zigzag representations of $\gl(1|1)$. We shall refer to them 
as (anti-)zigzag modules of $\sll(2|1)$, though based on the shape of their
(full) weight diagram it might be more appropriate to call them wedge 
modules. The (anti-)zigzag modules of $\sll(2|1)$ are parametrized by 
the number $d$ of their irreducible constituents and by the largest 
parameter $b\in\frac{1}{2}\Integer $ that appears among the atypical 
representations in their composition series. For our purposes it will
suffice to describe how (anti-)zigzag modules are built from atypical
representations   
\begin{equation} \label{zigzag} 
  \begin{split}
    \zig^d_\g(b): 
    &\ \ \bigoplus_{l=0}^{\lfloor\oh(d-1)\rfloor}\{b-l\}\lra
     \bigoplus_{l=\oh}^{\lfloor\oh d\rfloor-\oh}\{b-l\}\\[2mm]
    \azig^d_\g(b):
    &\ \ \bigoplus_{l=\oh}^{\lfloor\oh d\rfloor-\oh}\{b-l\}\lra
     \bigoplus_{l=0}^{\lfloor\oh(d-1)\rfloor}\{b-l\}\ \ .
  \end{split}
\end{equation}
Here, the symbol $\lfloor . \rfloor$ instructs us to take the integer 
part of the argument. Since we have simplified the diagrammatic 
presentation of the (anti-)\-zigzag modules in comparison to their 
counterparts for $\gl(1|1)$, we would like to stress that the 
structures are identical to the ones before. In particular, every 
invariant subspace $\{b'\}$ is a common submodule of both of its
neighbors $\{b'+\oh\}$ and $\{b'-\oh\}$ (should they be part of the 
composition series). Consequently, there exists the same dependence
on the parity of the parameter $d$. This also reflects itself in 
the behavior Kac modules under the action of the automorphism 
$\Omega$, 
\begin{align}
  \Omega\bigl(\zig^d_\g(b)\bigr)
  &\ =\ \begin{cases}
          \azig^d_\g(\frac{d-1}{2}-b) & \text{ for even }d\\[2mm]
          \zig^d_\g(\frac{d-1}{2}-b) & \text{ for odd }d\ \ . 
        \end{cases}
\end{align}
Similar formulas apply to (anti-)zigzag modules, only that all 
the $\zig$ must be replaced by $\azig$ and vice versa.  Let us 
finally point out that (anti-)Kac modules and atypical irreducible 
representations are just special cases of zigzag representations. 
The former correspond to the values $d=2$ and $d=1$ of the length 
$d$, respectively.%
\smallskip

This concludes our presentation of all finite dimensional representations
of $\sll(2|1)$. Throughout most of our discussion, we have not been 
very explicit, but in section \ref{sc:GLind} we shall see that many of 
the indecomposable representations of $\sll(2|1)$ may be induced 
from representations of $\gl(1|1)$. Along with our good insights 
into $\gl(1|1)$ modules, this then provides us with a rather 
direct construction of $\sll(2|1)$ representations. 

\begin{table}
  \centerline{\begin{tabular}{ccc}
    Symbol & Dimension & Type \\\hline\\
    $\{0\}=\zig^1_\g(0)=\azig^1_\g(0)$ & $1$ & atypical, irreducible \\[2mm]
    $\{j\}_\pm = \zig^1_\g(\pm j)= 
      \azig^1_\g(\pm j)$ & $4j+1$ & atypical, irreducible \\[2mm] 
    $\{b,j\} = \overline{\{b,j\}}$; $ b \neq \pm j$ & $8j$ & typical, 
      irreducible, projective\\[2mm] 
    $ \{\pm j,j\} = \zig^2_\g(\pm j), 
     \overline{\{\pm j,j\}} = \azig^2_\g(\pm j)$& $8j$ &
            indecomposable \\[2mm] 
    $\mc{P}_\g(0)$ & $8$ & indecomposable, projective\\[2mm]
    $\mc{P}_\g^\pm(j) = \mc{P}_\g(\pm j)$; $j>0$ & $16j+4$ & 
      indecomposable, projective\\[2mm]
    $\zig^d_\g(b)$, $\azig^d_\g(b)$ &  & indecomposable
  \end{tabular}}
  \caption{\label{tab:RepsSL}A complete list of finite dimensional 
  indecomposable representations of $\sll(2|1)$  (including 
  irreducibles) with diagonalizable 
  Cartan elements.}
\end{table}

\section{\label{sc:TP}Tensor products of $\sll(2|1)$ representations} 
\setcounter{equation}{0} 

In this section, we are going to address the main goal of this 
note, i.e.\ we shall determine all tensor products of finite 
dimensional $\sll(2|1)$ representations. Our results are partly
based on the previous analysis \cite{Marcu:1979sg} of certain 
special cases. The second important ingredient comes with our 
study of the $\gl(1|1)$ representation theory which enters through 
a particular embedding of $\gl(1|1)$ into $\sll(2|1)$. We
shall describe this embedding first before presenting our 
findings on the fusion of $\sll(2|1)$ representations.

\subsection{Decomposition with respect to $\gl(1|1)$}

Our main technical observation that will ultimately allow us 
to decompose arbitrary tensor products of finite dimensional 
$\sll(2|1)$ representations is a close correspondence with 
the representation theory of $\gl(1|1)$. The latter emerges 
from a particular embedding of $\gl(1|1)$ into $\sll(2|1)$. 
We shall specify this embedding in the first subsection. As 
an aside, we are then able to provide a much more explicit 
construction for many of the $\sll(2|1)$ representations we 
have introduced above. Finally, in the third subsection, we 
explain how finite dimensional representations of $\sll(2|1)$ 
decompose when restricted to $\gl(1|1)$.

\subsubsection{Embedding $\gl(1|1)$ into $\sll(2|1)$}

In order to embed the Lie superalgebra $\gl(1|1)$ into $\sll(2|1)$ 
we shall employ the following regular monomorphism $\epsilon$, 
\begin{align}
  \epsilon(E)&\ =\ Z-H&
  \epsilon(N)&\ =\ Z+H&
  \epsilon(\psi^+)&\ =\ F^+&
  \epsilon(\psi^-)&\ =\ \bar{F}^-\ \ . 
\end{align}
There exist different embeddings which arise by concatening $\epsilon$ 
with $\Omega$ and/or $\omega$ but we will not consider them since 
apparently they do not give rise to any new information. Let us 
point out, though, that $\epsilon$ does not intertwine the actions 
of the outer automorphism $\omega$ and $\Omega$, i.e.\ $\Omega \circ 
\epsilon \neq \epsilon \circ \omega$.

\subsubsection{\label{sc:GLind}Induced representations from $\gl(1|1)$}

As we have anticipated, we can exploit the relation between $\gl(1|1)$ 
and $\sll(2|1)$ to construct representations of the latter from the 
former. To this end, we note that the embedding of $\gl(1|1)$ induces 
the following decomposition of $\sll(2|1)$ into eigenspaces of the 
element $\epsilon(E)$,
\begin{equation}
  \mf{g}\ =\ \mf{k}_1\oplus\mf{k}_0\oplus\mf{k}_{-1}\ \ ,
\end{equation}
where $\mf{k}_0=\gl(1|1)$, $\mf{k}_1=\Span\{E^+,\bar{F}^+\}$ and
$\mf{k}_{-1}=\Span\{E^-,F^-\}$ such that $[\mf{k}_i,\mf{k}_j]\subset
\mf{k}_{i+j}$. Given any representation $\rho_\h$ of $\gl(1|1)$ we can 
thus induce a module of $\sll(2|1)$ using the elements of $\mf{k}_1$ 
(or $\mf{k}_{-1}$) as generators. The resulting representation is 
infinite dimensional but under certain circumstances one may take a 
quotient and end up with a finite dimensional representation space. 
A condition in the choice of the $\gl(1|1)$ representation $\rho_\h$ 
arises in particular from considering the $\sll(2)$ multiplets 
within the induced representation. In order for the latter to 
possess a finite dimensional quotient, the spectrum of the 
Cartan element $2H$ must by integer. Since $2H$ is the image 
of $N-E$ under the monomorphism $\epsilon$, we conclude that 
$\rho_\h$ is only admissible if $\rho_\h(N-E)$ has integer spectrum. In 
the case of a typical representation $\rho_\h = \langle e,n\rangle$, 
for example, our condition restricts $e-n$ to be an integer. 
\smallskip

Many $\sll(2|1)$-representations can actually be obtained through 
such an induction. This applies in particular to the projective 
covers $\mc{P}_\g^\pm(j)$ with $j\neq 0$ which are obtained from 
$\rho_\h = \P_\h(\pm 2j)$. In the case of the (anti-)zigzag  modules 
$\zig^d_\g(b)$ and $\azig^d_\g(b)$, we only need to avoid the 
range $0 < 2b < d-1$. Outside this interval, we can obtain the 
(anti-)zigzag representations by induction, using the $\gl(1|1)$
representations $\zig^d_\h(2b)$ and $\azig^d_\h(2b)$ for $\rho_\h$.   
\smallskip

What makes the induction particularly interesting for us is another
aspect: Suppose we start with a $\gl(1|1)$-representation $\rho_\h$ 
in which $\rho_\h(E) = 0$. Since $[\epsilon(E),\mf{k}_{\pm 
1}] = \pm \mf{k}_{\pm 1}$, our creation operators cannot generate 
any additional eigenstates of $\rho_\h(E)$ with vanishing eigenvalue. 
In other words, if $\rho_\g$ is an $\sll(2|1)$ representation 
which can be obtained by our induction from $\rho_\h$ and if 
$\rho_\h(E) = 0$, then the decomposition of $\rho_\g$ into 
representations of $\h$ can only contain typical representations
in addition to the representation $\rho_\h$ we started with. 
We shall find that this observations extends to a simple 
correspondence between atypical representations (and their 
indecomposable composites) of $\sll(2|1)$ and $\gl(1|1)$.

\subsubsection{Decomposition of $\sll(2|1)$ representations}

Before we decompose representations of $\g$ into representations
of $\h$ we introduce a few new notations that will become quite 
useful. In particular, we will employ a map $\mc{E}$ which takes  
irreducible representations of $\g$ and turns them into a very 
specific sum of typical $\h$ representations. On atypical 
representations, $\mc{E}$ is defined by 
\begin{equation}
   \mc{E}\bigl(\{j\}_\pm\bigr)\ =\ \bigoplus_{n=1}^{2j}\,\langle\pm n, 
        \oh\pm(2j+\oh-n)\rangle\ \ .
\end{equation}
We shall claim below that $\mc{E}(\{j\}_\pm)$ contains all the 
typical $\gl(1|1)$-representations that appear in the decomposition 
of $\{j\}_\pm$. Similarly, we may define 
\begin{equation}
   \mc{E}\bigl(\{b,j\}\bigr)\ =\ \bigoplus_{n=-j+1}^{j}\!\!\!\!'\ \ 
      \Bigl[\langle b+1-n,
          b+n\rangle\oplus\langle b-n,b+n\rangle\Bigr]
\end{equation}
on typical representations $\{b,j\}, b \neq \pm j$. The prime $'$ on 
the summation symbol instructs us to omit all terms that correspond to 
atypical representations. We can extend $\mc{E}$ linearly to all 
completely reducible representations of $\sll(2|1)$. 
\smallskip 

Another map $\mc{S}_\g$ converts indecomposable representations of 
$\sll(2|1)$ into semi-simple modules, namely into the sum of all
irreducible representations that appear in the decomposition series. 
Explicitly, we have 
\begin{equation}
  \begin{split}
\mc{S}_\g\bigl(\P_\g(j)\bigr) &\ =\ 2\{j\}\oplus \{j-\sfrac12\} \oplus
        \{j+\sfrac12\} \ \ , \label{S}\\[2mm] 
\mc{S}_\g\bigl(\zig^d_\g(b)\bigr) &\ =\ \bigoplus_{l=0}^{d-1} \ 
\{b-\frac{l}{2}\} 
\ \ . 
  \end{split}
\end{equation} 
The expressions should be compared with our diagrams \eqref{P} and 
\eqref{zigzag} for the projective covers and the (anti-) zigzag 
modules of $\sll(2|1)$.  
\smallskip 

Once this notation is introduced, our decomposition formulas take a 
particularly simple form. For the atypical representations and their 
composites one obtains 
\begin{equation}
  \begin{split}
  \{j\}\bigr|_\h  &\ =\   
 \langle 2j\rangle \  \oplus\ \mc{E}\bigl( \{j\}\bigr) \\[2mm]  
  \mc{P}_\g (j) \bigr|_{\h}
  &\ = \ \mc{P}_\h(2j) \ \oplus \ \mc{E} \circ 
       \mc{S}_g\bigl(\P_\g(j)\bigr) \label{ghdec} \\[2mm] 
    \zig^d_\g(b)\bigr|_{\h}
    &\ = \ \zig^d_\h(2b)\ \oplus\ \mc{E} \circ 
    \mc{S}_\g\bigl(\zig^d_\g(b)\bigr) \ \ .
  \end{split}
\end{equation}
The last relation also holds for anti-zigzag modules if we replace
all $\zig$ by $\azig$. Note that, up to typical contributions, there 
is a one-to-one correspondence between the $\sll(2|1)$ representations
on the left and the $\gl(1|1)$ representations on the right hand side. 
Things are slightly more complicated for the typical representations of 
$\sll(2|1)$ for which the decomposition is given by  
\begin{equation}
  \{b,j\}\bigr|_{\h}
  \ =\ \begin{cases}
          \ \mc{E}\bigl(\{b,j\}\bigr) & \text{ for } 
           \ \ b \ \neq \ -j, \dots, j\ , \\[2mm]
\ \P_\h (2b)\, \oplus \ \mc{E}\bigl(\{b,j\}\bigr) 
 & \text{ for }  \ \ b \ = \ -j+1, \dots, j-1\ .\end{cases}  
\end{equation}
\smallskip
Note that in the second case, the image of the symbol $\mc{E}$ 
contains only $4j-2$ typical representations so that the 
dimensions match.

\subsection{Decomposition of $\sll(2|1)$ tensor products}

We are finally prepared to decompose arbitrary tensor products 
of finite dimensional $\sll(2|1)$ representations. Our presentation 
below is split into three different parts. We shall begin by reviewing 
Marcu's results \cite{Marcu:1979sg} on the decomposition of tensor 
products between two typical representations and between a typical and 
an atypical representation. The extension to arbitrary tensor products 
involving one typical representation is then straightforward. 
The second subsections contains new results on tensor products
in which at least one factor is a projective cover. Finally, we
shall decompose arbitrary tensor products of two (anti-)zigzag 
modules.  

\subsubsection{Tensor products involving a typical representation}  

Before presenting Marcu's results, we would like to introduce 
some notation that will permit us to rephrase the answers in a
much more compact form. To this end, let us define a map $\pi$ 
which sends representations of the bosonic subalgebra $\g^{(0)}$ 
to typical representations of $\g$. Its action on irreducibles
is given by 
\begin{equation}
\pi(b-\sfrac12,j-\sfrac12) \ = \ \left\{ 
 \begin{array}{ll} \{b,j\} & \text{ for } \ \ b \neq \pm j \ \ , 
                   \\[2mm] 
                   0 & \text{ for } \ \ b = \pm j \ \ .  
\end{array} \right. 
\end{equation} 
The map $\pi$ may be extended to a linear map on the space of 
all finite dimensional representations of $\g^{(0)}$. 
\smallskip 

The first tensor product we would like to display is the one 
between two typical representations \cite{Marcu:1979sg}. In 
our new notations, the decomposition is given by 
\begin{eqnarray} 
 \{b_1,j_1\}\otimes\{b_2,j_2\} & = &  
   \pi\bigl((b_1-\oh,j_1-\oh)\otimes {\{b_2,j_2\}\bigr|_{\g^{(0)}}}\bigr)
      \  \oplus\ \\[4mm] \nonumber & & \hspace*{-4cm} \oplus \  
    \left\{ \begin{array}{cl} 
   \mc{P}_\g (\pm |b_1+b_2|\mp \oh) &  \text{for} \  
                  b_1+b_2\ =\ \pm(j_1+j_2) \\[2mm] 
\mc{P}_\g^\pm(|b_1+b_2|)\oplus\mc{P}_\g^\pm(|b_1+b_2|-\oh)
  & \text{for} \ b_1+b_2\in \pm
  \{|j_1-j_2|+1,\dots,j_1+j_2-1\} \\[2mm] 
      \mc{P}_\g(\pm |b_1+b_2|) & \text{for} \  
    b_1+b_2 \ = \ \pm |j_1-j_2|\ \ .  
\end{array} \right. 
\end{eqnarray}    
Note that neither $j_1$ nor $j_2$ can vanish so that the three 
cases listed above are mutually exclusive. If none of them applies, 
the tensor product contains only typical representations. These are 
computed by the first term. All it requires is the decomposition of typical 
$\g$ representations into irreducibles of the bosonic subalgebra 
(see eq.\ \eqref{typdec}) and a computation of tensor products for 
representations of $\g^{(0)} = \gl(1) \oplus \sll(2)$ which presents 
no difficulty. The outcome is then converted into a direct sum of 
typical representations through our map $\pi$.  
\smallskip 

Tensor products of typical with atypical representations can also
be found in Marcu's paper. The results are 
\begin{eqnarray} 
 \{b_1,j_1\}\otimes\{j_2\}_\pm & = &  
   \pi\bigl( (b_1-\oh,j_1-\oh)\otimes \{j_2\}_\pm\bigr|_{\g^{(0)}} \bigr)
      \  \oplus\ \label{eq:typatyp}\\[4mm] \nonumber & & \hspace*{-2cm} \oplus \  
    \left\{ \begin{array}{cl} 
   \mc{P}^\mp_\g (|b_1\pm j_2|- \oh) & \ \ \ \ \text{for} \ \  
                  b_1\pm j_2\ \in \ \mp \{|j_1-j_2|+1, \dots, j_1+j_2 \} 
    \\[2mm] 
      \mc{P}^\pm_\g(|b_1\pm j_2|) & \ \ \ \ \text{for} \ \  
    b_1\pm j_2 \ \in \ \pm \{|j_1-j_2|, \dots, j_1+j_2-1\} \ .  
\end{array} \right. 
\end{eqnarray}    
This formula can also be used to determine the tensor product of 
typical representations with any composite of atypical representations, 
i.e.\ with the projective covers and the (anti-)zigzag modules. In fact, 
these tensor products are simply given by 
\begin{equation} \label{TX} 
\{b,j\} \otimes \mc{H} \ = \ \{b,j\} \otimes 
          \S_\g(\mc{H}) \ \ \ \ \text{ for } \ \ \ 
    \mc{H} \ = \ \P_\g(l), \zig^d_\g(l) \text{ or } 
 \azig^d_\g(l) \ \ .   
\end{equation} 
Such an outcome is natural since the decomposition of a tensor 
product of a typical with any other representation is known to 
be decomposable into typicals and projective covers. One may 
determine the exact content through the $\gl(1|1)$ embedding
and it is rather easy to see that the answers may always be
reduced to the computation of tensor products with atypical 
irreducibles, as it is claimed in equation \eqref{TX}.   

\subsubsection{Tensor products involving a projective cover} 

This subsection collects all our findings on tensor products 
involving at least one projective cover. General results 
guarantee that such tensor products decompose into a sum of 
projective representations. The result for the tensor product 
of a projective cover with a typical representation has been 
spelled out already (see eq.\ \eqref{TX}). Therefore, we can 
turn directly to the next case, the product of two projective 
covers. 
\bigskip 

\noindent 
{\bf Proposition 1:} {\em The tensor product between two 
projective covers $\P^\pm_\g(j_1), j_1 \geq 0,$ and $\P_\g(j_2) 
= \P^{\sign(j_2)}(|j_2|)$ is given by  
\begin{eqnarray} 
 \P^\pm_\g(j_1) \otimes \P_g(j_2) & = &  
   \pi\bigl(\ H^\pm_{j_1}  
        \otimes \P_\g(j_2)\bigr|_{\g^{(0)}}\bigr)
      \  \oplus\ \\[4mm] \nonumber & & 
\hspace*{-1cm} \oplus \  
\P_\g(\pm j_1+ j_2+\oh) \oplus 2\cdot \P_\g(\pm j_1+ j_2) \oplus 
\P_\g(\pm j_1+ j_2-\oh) \ \\[2mm] 
\text{ where }  \ \ \ H^\pm_j & = & 
(\pm j -\oh,j-\oh) \oplus (\pm (j +\oh)-\oh,j) \ \ \ \ \text{for } 
 \ \ j > 0  
\end{eqnarray}    
and $H_0 = H_0^\pm = (0,0)\oplus(-1,0)$. 
In the argument of $\pi$ the product $\otimes$ refers to the 
fusion between representations of the bosonic subalgebra 
$\g^{(0)} = \gl(1) \oplus \sll(2)$.}  
\medskip 

\begin{proof}[\sc Proof:]
Our claim concerning typical representations
in the decomposition requires little comment. Let us only 
stress that the two bosonic multiplets $(\pm (j+\oh)-\oh,j)$ and 
$(\pm j - \oh,j-\oh)$ that appear in the space $H_j^\pm$ 
are the ground states of the two Kac modules from 
which $\P^\pm_\g(j)$ is composed (see eq.\ \eqref{P}). The 
contributions from projective covers, on the other hand, may 
be deduced from the embedding of $\gl(1|1)$ along with the
formula \eqref{hPP} for tensor products of the projective 
covers $\P_\h$.
\end{proof} 
\medskip 

\noindent 
{\bf Proposition 2:} {\em The tensor product between a
projective cover $\P^\pm_\g(j), j \geq 0,$ and a zigzag 
module $\zig^d_\g(b)$ is given by  
\begin{eqnarray} \nonumber 
 \P^\pm_\g(j) \otimes \zig^d_\g(b) & = &  
   \pi\bigl(\ H^\pm_{j} \otimes 
     \zig^d_\g(b)\bigr|_{\g^{(0)}}\bigr)
\ \oplus \ \bigoplus_{p=0}^{d-1} \ \P_\g(\pm j + b - \oh p) 
\end{eqnarray}    
where $H^\pm_j$ is the same as in proposition 1. 
To determine the tensor product with an anti-zigzag module
$\azig^d_\g(b)$, we replace $\zig_\g$ by $\azig_\g$.}  
\bigskip 

\begin{proof}[\sc Proof:]
The statement is established in the same way as 
proposition 1, using formula \eqref{hPZ} as input from the 
representation theory of $\gl(1|1)$.
\end{proof}

\subsubsection{Tensor products between (anti-)zigzag modules}

In the following we shall denote the fusion ring of finite dimensional 
representations of a Lie superalgebra $\g$ by $\mf{Rep}(\g)$. As we 
remarked before, projective representations of $\g$ form an ideal in 
$\mf{Rep}(\g)$. The latter will be denoted by $\mf{Proj}(\g)$. Our 
results on the decomposition of $\sll(2|1)$ representations into 
representations of $\gl(1|1)$ imply the following nice result. 
\medskip 

\noindent
{\bf Proposition 3:} {\em Modulo projectives, the representation ring
of $\g = \sll(2|1)$ may be embedded into the representation ring of
$\h=\gl(1|1)$, i.e.\ there exists a monomorphim $\vartheta$, 
$$ \vartheta: \mf{Rep}(\g)/\mf{Proj}(\g)\  \longrightarrow
         \   \mf{Rep}(\h)/\mf{Proj}(\h) \ \ . $$ 
Note that $\mf{Rep}(\g)/\mf{Proj}(\g)$ is generated by (anti-)zigzag 
modules. On the latter, the mono\-morphism $\vartheta$ acts 
according to 
$$ \vartheta\bigl(\zig_\g^d(b)\bigr)\ = \ \zig_\h^d(2 b) \ \ \ , \ \ \ 
       \vartheta\bigl(\azig_\g^d(b)\bigr)\ = \ \azig_\h^d(2 b)\ \ . $$}%
\begin{proof}[\sc Proof:]
This proposition is an obvious consequence of the formulas 
\eqref{ghdec} for the decomposition of $\sll(2|1)$ representations into 
indecomposables of $\gl(1|1)$.
\end{proof} 
\bigskip

This proposition can be used to compute the non-projective contributions
of tensor products between (anti-)zigzag representations explicitly from 
our $\gl(1|1)$ formulas. For the tensor product of two atypical 
representations one finds in particular 
$$ \{j_1\} \otimes \{j_2\} \ = \ \{j_1+j_2\} \ 
               \text{mod} \ \mf{Proj}\bigl(\sll(2|1)\bigr)\ \ . 
$$ 
The answer is in agreement with the findings of Marcu who has computed 
the tensor product of atypical representation in \cite{Marcu:1979sg}. 
In fact, the full answer for the tensor product of two atypical 
representations is encoded in the formulas 
\begin{eqnarray} 
  \{j_1\}_\pm\otimes\{j_2\}_\pm
  & = &  \bigl\{j_1+j_2\bigr\}_\pm\ \oplus\ 
 \bigoplus_{j=|j_1-j_2|}^{j_1+j_2-1}\{\pm(j_1{+}j_2{+}\oh),j+\oh\}\ \ , 
\\[2mm] 
\{j_1\}_+\otimes\{j_2\}_-
  & = &  \bigl\{|j_1-j_2|\bigr\}_{\sign(j_1-j_2)}\ \oplus\ 
  \bigoplus_{j=|j_1-j_2|+1}^{j_1+j_2}\{j_1{-}j_2,j\}\ \ .\label{eq:atypatyp}
\end{eqnarray} 
Let us agree to denote the sums of typical representations that 
appear on the right hand side by $\mc{T}( \{j_1\}_\pm,\{j_2\}_\pm)$ 
and $\mc{T}(\{j_1\}_+,\{j_2\}_-)$, respectively. Furthermore, we 
would like to extend $\mc{T}$ to a bi-linear map on arbitrary sums
of atypical irreducibles. The map $\mc{T}$ features in the following 
decomposition of tensor products between two (anti-)zigzag modules. 
\bigskip 

\noindent
{\bf Proposition 4:} {\em Tensor product between two zigzag 
modules of $\sll(2|1)$ can be decomposed as follows 
\begin{equation}\label{ZZdec}  
 \zig^{d_1}_\g(b_1) \otimes \zig^{d_2}_\g(b_2)\ = \ 
   \mc{T}(\S_\g(\zig^{d_1}_\g(b_1)), \S_\g(\zig^{d_2}(b_2)) \ 
   \oplus \ \Theta(\zig^{d_1}_\h(2b_1) \otimes \zig^{d_2}_\h(2b_2)) 
 \ \ . 
\end{equation}  
The map $\mc{T}$ was introduced in the text preceding this 
proposition and $\S_g$ replaces its argument by a direct 
sum of irreducibles in the decomposition series (see eqs.\ 
(\ref{S})). $\Theta$ is a linear map that replaces 
certain $\h$-representations by $\g$-representations according to 
$$ \Theta(\P^d_\h(n)) = \P^d_\g(\oh n) \ \ \ , \ \ \ 
   \Theta(\zig^d_\h(n)) = \zig^d_\g(\oh n) \ \ \ , \ \ 
  \Theta(\azig^d_\h(n)) = \azig^d_\g(\oh n)\ \ . 
$$ 
Analogous formulas apply to tensor product of zigzag with 
anti-zigzag modules and to the fusion of two anti-zigzag 
representations.} 
\bigskip 

\begin{proof}[\sc Proof:]
The rule that determines the contribution from typical 
representations is fairly obvious and the (anti-)zigzag representations
in the tensor product are a consequence of proposition 3. The terms
involving projective covers, finally, can be found through the 
decomposition into $\gl(1|1)$ representations. This part is the 
most subtle, since projective $\gl(1|1)$ representations can in 
principle arise through the decomposition of both projective covers 
and typical $\sll(2|1)$ representations. To see that projective 
covers for $\sll(2|1)$ representations contribute to the 
decomposition only through the second term, we note that all 
the atypical components that appear in the tensor product of 
the $\sll(2|1)$ zigzag representations are needed to build 
the image of $\Theta$ on the right hand side of eq.\ 
\eqref{ZZdec}. Hence, all projective covers of 
$\gl(1|1)$ atypicals that are not found in the restriction 
of $\Theta(\zig^{d_1}_\h(2b_1) \otimes \zig^{d_2}_\h(2b_2))$
must arise from a restriction of typical $\sll(2|1)$ 
representations. 
\end{proof}

\bigskip 
\bigskip
\bigskip
\noindent{\bf Acknowledgements:} It is a pleasure to thank
  Jerome Germoni, Hubert Saleur, Paul Sorba and Anne Taormina for
  many useful discussions. This work was partially supported by the
  EU Research Training Network grants ``Euclid'', contract number
  HPRN-CT-2002-00325, ``Superstring Theory", contract number
  MRTN-CT-2004-512194, and ``ForcesUniverse'', 
  contract number MRTN-CT-2004-005104.
  TQ is supported by a PPARC postdoctoral fellowship under reference
  PPA/P/S/2002/00370 and partially by the PPARC rolling grant
  PPA/G/O/2002/00475. We are grateful for the kind hospitality at the
  ESI during the workshop ``String theory in curved backgrounds'' which
  stimulated the present work.%


\def\cprime{$'$} \def\cprime{$'$}
\providecommand{\href}[2]{#2}\begingroup\raggedright\endgroup

\end{document}